\documentstyle[referee,psfig]{mn}

\newif\ifAMStwofonts

\ifoldfss
  \ifCUPmtlplainloaded \else
    \NewTextAlphabet{textbfit} {cmbxti10} {}
    \NewTextAlphabet{textbfss} {cmssbx10} {}
    \NewMathAlphabet{mathbfit} {cmbxti10} {} % for math mode
    \NewMathAlphabet{mathbfss} {cmssbx10} {} %  "   "    "
  \fi
  \ifAMStwofonts
    \ifCUPmtlplainloaded \else
      \NewSymbolFont{upmath} {eurm10}
      \NewSymbolFont{AMSa} {msam10}
      \NewMathSymbol{\upi}     {0}{upmath}{19}
      \NewMathSymbol{\umu}     {0}{upmath}{16}
      \NewMathSymbol{\upartial}{0}{upmath}{40}
      \NewMathSymbol{\leqslant}{3}{AMSa}{36}
      \NewMathSymbol{\geqslant}{3}{AMSa}{3E}

    \fi
  \fi
\fi % End of OFSS

\ifnfssone
  \newmathalphabet{\mathit}
  \addtoversion{normal}{\mathit}{cmr}{m}{it}
  \addtoversion{bold}{\mathit}{cmr}{bx}{it}
  \newmathalphabet{\mathbfit} % math mode version of \textbfit{..}
  \addtoversion{normal}{\mathbfit}{cmr}{bx}{it}
  \addtoversion{bold}{\mathbfit}{cmr}{bx}{it}
  \newmathalphabet{\mathbfss} % math mode version of \textbfss{..}
  \addtoversion{normal}{\mathbfss}{cmss}{bx}{n}
  \addtoversion{bold}{\mathbfss}{cmss}{bx}{n}
  \ifAMStwofonts
    \ifCUPmtlplainloaded \else
      \UseAMStwoboldmath
      \makeatletter
      \new@mathgroup\upmath@group
      \define@mathgroup\mv@normal\upmath@group{eur}{m}{n}
      \define@mathgroup\mv@bold\upmath@group{eur}{b}{n}
      \edef\UPM{\hexnumber\upmath@group}
      \new@mathgroup\amsa@group
      \define@mathgroup\mv@normal\amsa@group{msa}{m}{n}
      \define@mathgroup\mv@bold\amsa@group{msa}{m}{n}
      \edef\AMSa{\hexnumber\amsa@group}
      \makeatother
      \mathchardef\upi="0\UPM19
      \mathchardef\umu="0\UPM16
      \mathchardef\upartial="0\UPM40
      \mathchardef\leqslant="3\AMSa36
      \mathchardef\geqslant="3\AMSa3E
    \fi
  \fi
\fi % End of NFSS release 1
\ifnfsstwo
  \DeclareMathAlphabet{\mathbfit}{OT1}{cmr}{bx}{it}
  \SetMathAlphabet\mathbfit{bold}{OT1}{cmr}{bx}{it}
  \DeclareMathAlphabet{\mathbfss}{OT1}{cmss}{bx}{n}
  \SetMathAlphabet\mathbfss{bold}{OT1}{cmss}{bx}{n}
  \ifAMStwofonts
    \ifCUPmtlplainloaded \else
      \DeclareSymbolFont{UPM}{U}{eur}{m}{n}
      \SetSymbolFont{UPM}{bold}{U}{eur}{b}{n}
      \DeclareSymbolFont{AMSa}{U}{msa}{m}{n}
      \DeclareMathSymbol{\upi}{0}{UPM}{"19}
      \DeclareMathSymbol{\umu}{0}{UPM}{"16}
      \DeclareMathSymbol{\upartial}{0}{UPM}{"40}
      \DeclareMathSymbol{\leqslant}{3}{AMSa}{"36}
      \DeclareMathSymbol{\geqslant}{3}{AMSa}{"3E}
    \fi
  \fi
\fi % End of NFSS release 2

\ifCUPmtlplainloaded \else
  \ifAMStwofonts \else % If no AMS fonts
    \def\upi{\pi}
    \def\umu{\mu}
    \def\upartial{\partial}
  \fi
\fi
\title[4U\,1630$-$47]
  {Complex outburst behaviour from the black-hole candidate 4U\,1630$-$47}

\author[Kuulkers et al.]
  {E.~Kuulkers$^1$\thanks{E-mail: e.kuulkers1@physics.oxford.ac.uk (EK),
  aparmar@astro.estec.esa.nl (ANP), kitamoto@ess.sci.osaka-u.ac.jp (SK),
  lynnc@charmian.sonoma.edu (LRC), r-sood@adfa.edu.au (RKS).}
  A.~N.~Parmar,$^{2\star}$ S.~Kitamoto,$^{3\star}$ L.~R.~Cominsky$^{4\star}$ and
  \newauthor
  R.~K.~Sood$^{5\star}$\\
  $^1$Astrophysics, University of Oxford, Nuclear and Astrophysics Laboratory, 
  Keble Road, Oxford OX1 3RH, United Kingdom\\
  $^2$ESA/ESTEC, Astrophysics Division, P.O.~Box 299, 2200~AG Noordwijk, 
  The Netherlands\\
  $^3$Department of Earth and Space Science, Graduate School of Science,
  Osaka University, 1-1, Machikaneyama-cho, Toyonaka,\\ Osaka 560, Japan\\
  $^4$Sonoma Sate University, Physics and Astronomy Department,
  1801 E.~Cotati Avenue, Rohnert Park, CA 94928-3609, USA\\
  $^5$Physics Department, University College, University of New South
  Wales, Canberra ACT 2600, Australia (UNSW)}

\date{Accepted. Received 1997 January 15}
\pagerange{\pageref{firstpage}--\pageref{lastpage}}
\pubyear{1997}

\begin{document}

\label{firstpage}

\maketitle

\begin{abstract}

We present data from different epochs in 1978, 1987--1991 and 1996 
from the black-hole candidate 4U\,1630$-$47.
For the first time we present almost complete coverage of the 
outbursts which started in 1987, 1988, and 1996.
We find that the outburst behaviour of 4U\,1630$-$47 is more complex than
previously realized. The source shows outbursts with durations on the order of
$\sim$100-200 days and sometimes intervals of long-term X-ray activity. 

The relatively short outbursts which occurred in 1987 and 1996 exhibited 
different colour behaviour: the outburst in 1987 showed softening 
of the X-ray spectrum, whereas the outburst in 1996 showed 
hardening of the X-ray spectrum, as the outbursts proceeded.

The outburst which started in 1977 may have lasted for up to $\sim$10~months, 
whereas the outburst which started in 1988 showed activity for up to 
$\sim$2.4~years. Such long-term activity is reminiscent of that seen in 
GRS\,1716$-$249 and in the galactic superluminal sources GRS\,1915+105 and 
GRO\,J1655$-$40.

We refine the outburst ephemeris of 4U\,1630$-$47 and find that 
the outburst recurrence time scale may have changed from 
$\sim$600~days to $\sim$690~days between the 1984 and 1987 outbursts.
If the recurrence interval
of $\sim$690~days continues, the next outburst of 4U\,1630$-$47 is predicted
to occur in 1998 January.

\end{abstract}


\section{Introduction}

Soft X-ray transients (SXTs) are low-mass X-ray binaries (LMXBs)
which appear suddenly in the X-ray sky and generally decay within weeks to 
several months. Between outburst most systems
remain quiescent for up to several decades
(see, e.g.\ Chen, Shrader \&\ Livio 1997, and references therein).
The X-ray outburst light curves of SXTs show a wide variety of shapes,
such as a fast rise and exponential decay (so-called `FRED' light curve),
flat topped or plateau, or irregularly shaped
(see White, Kaluzienski \&\ Swank 1984, Chen et al.\ 1997).

The compact member of an SXT is
thought to be either a neutron star or a black hole. Although various
X-ray characteristics have been put forward to
distinguish binaries containing a black hole from those containing a neutron 
star, the situation is unclear, since many characteristics once thought to be 
indicative of a black hole are also found in neutron star binaries, or are
absent in black-hole binaries whose masses have been dynamically determined 
(see e.g.\ van der Klis 1994, 1995, Tanaka \&\ Lewin 1995, 
Tanaka \&\ Shibazaki 1996).
These characteristics include a strong
ultra-soft component with a high-energy power-law tail near maximum 
X-ray intensities, or strong flickering at low X-ray 
intensities. The best indication for the presence of a black hole comes from 
dynamical measurements, where the inferred primary mass clearly exceeds 
the maximum mass of a neutron star (see, e.g.\ Charles 1996, Casares 
1997, for recent reviews).

The SXT 4U\,1630$-$47 is one of these transients suspected to 
harbor a black hole based on the X-ray spectral and timing behaviour
(White et al.\ 1984; Parmar, Stella \&\ White 1986; 
\mbox{Kuulkers}, van der Klis \&\ Parmar 1997). 
No likely optical counterpart has been reported, probably due to reddening 
and the crowded field within the 
X-ray uncertainty region (see Parmar et al.\ 1986).
4U\,1630$-$47 is the shortest known recurrent black-hole SXT (BHXT), with
an outburst recurrence interval of $\sim$600 days (Jones et al.\ 1976, 
Priedhorsky 1986, Parmar, Angelini \&\ White 1995,
Parmar et al.\ 1997). The outbursts are not strictly periodic,
varying by $\sim$8\%\ rms (Parmar et al.\ 1997). 

An extreme example of the deviation from the $\sim$600 days recurrence
interval is the outburst which started in November 1977, $\sim$70~days
later (Kaluzienski \&\ Holt 1977) than predicted 
(Grindlay 1977) from the previous outburst. 
Observations about a month earlier showed that 4U\,1630$-$47 was below 
the detection limits of {\it HEAO}-1 (Share et al.\ 1978, 
Kaluzienski et al.\ 1978). 
We note that the deviation of the 1977 outburst time from the outburst 
ephemeris is 3--4~months (Parmar et al.\ 1995, 1997).
Observations $\sim$4~months and $\sim$6~months after the start of this 
outburst by {\it HEAO}\,A-1 (Share et al.\ 1978) and {\it HEAO}\,A-2 
(Kaluzienski et al.\ 1978), and Ariel\,V (Sims \&\ Watson 1978), 
respectively, revealed that the source was still in outburst.

We present additional observations by {\it HEAO}\,A-1 in 1978 September, 
during which period the source was clearly detected. This suggests that the 
source was active for at least 10~months. Data from the {\it Ginga} All Sky 
Monitor (ASM) confirm the existence of such long-term activity, which is
also reported here. The outburst from 4U\,1630$-$47 which occurred in 1996 was 
detected by {\it RXTE} (Marshall 1996; Levine et al.\ 1996b).
We show this outburst was a relatively short one, and only 
lasted for $\sim$4~months.

\section{Observations and analysis}

The {\it HEAO}\,A-1 data analyzed here were
obtained during 1978 March 1--8 and 1978 September 
2--8 with the scanning module detectors (see Wood et al.\ 1984
for further experimental details). We present data from module~3. 
Module~3 had an effective area of 1650\,cm$^2$. The instrument was sensitive 
in the 0.25--25\,keV energy range. After eliminating 
data that were Earth-occulted or contaminated by the South 
Atlantic Anomaly region of high-charged particle background, 
the count rates of the remaining "good" observations of 4U1630$-$47 
were fitted by a multi-parameter model. This model typically fitted 15
parameters to a 500 data point sample of scanning data
centered on the expected position of 4U\,1630$-$47.  Two
parameters were used for a linear fit to the background;
the X-ray count rates for each of 13 known nearby sources in the field 
of view were also fitted. Most of the sources had zero 
count rates on a typical scan. After fitting, the count rates and errors 
were corrected for the collimator transmission. 
These data were previously reported by Cominsky et al.\ (1994).

The ASM onboard the {\it Ginga} satellite viewed the
X-ray sky between 1987 February to 1991 October. Its effective
area was 420\,cm$^2$, with three criss-crossed $45\degr \times 1\degr$
full-width-at-half-maximum (FWHM) fan-beam collimators. For a more detailed 
account on the {\it Ginga} ASM we refer to Tsunemi et al.\ (1989).  
The ASM scanned the sky at intervals of typically a few days, 
when the satellite was rotated around the $z$-axis for about 20\,min.
During each scan across a source, 16-channel 1--20\,keV source spectra 
were obtained. Typical exposure times were $\sim$3--18\,s,
depending on the source latitude in the spacecraft equatorial ($x-y$)
plane. The 5$\sigma$ confidence detection limit was about 
50\,mCrab (1--6 keV), for favorably located sources. This limit increased
for sources located further from the spacecraft equatorial plane.

We selected {\it Ginga} ASM data which had a low and stable background, where 
the source was not occulted by the Earth, and whenever the spacecraft
aspect was acceptable (i.e.\ when the source was within 25$\degr$ of the 
center of the {\it Ginga} ASM's field of view).  
This procedure resulted in a total of 220 4U\,1630$-$47 measurements
during the 4.5~year {\it Ginga} mission.
In this paper {\it Ginga} ASM colours are defined as the ratio of the 
source count rates in the 6--20\,keV and 1--6\,keV energy bands.
We present colour values whenever the 1--20\,keV source count rate 
exceeded 0.2\,cts\,s$^{-1}$\,cm$^{-2}$.

The ASM onboard the Rossi X-ray Timing Explorer ({\it RXTE})
consists of three Scanning Shadow Cameras (SCCs) mounted on a rotating drive
assembly. Each camera includes a position sensitive proportional 
counter, has a FWHM field of view of $6\degr \times 90\degr$, and
a geometric area of $\sim$30\,cm$^2$.
The {\it RXTE} ASM scans the X-ray sky in series of $\sim$90\,s dwells in 
three energy bands, corresponding approximately to
1.5--3\,keV, 3--5\,keV and 5--12\,keV. Due to satellite motion and a 40\%\ 
duty cycle, any given source is scanned 5--10 times 
per day. For a more detailed discussion of the {\it RXTE} ASM see Levine et 
al.\ (1996a).

We used the 1-dwell data of the {\it RXTE} ASM products, which were made 
available through the {\it High Energy Astrophysics Science Archive Research 
Center} (HEASARC). The source count rates were obtained by fitting the 
observed count rates in the field of view, using the positions of the known 
and detected sources collected in the {\it RXTE} ASM source catalogue. This 
fitting was performed for all three SSCs individually, as well as for the 
three energy bands. In our analysis we only used data of the total count rate 
of each SSC and the count rates of each SSC in the three energy bands 
which yielded $\chi^2_{\rm red}<1.5$ from the fits. 
Data points of SSC1 and SSC2 which were taken simultaneously,
were averaged. We then computed the daily weighted averages.
In this paper the {\it RXTE} ASM colours are defined as the ratio of the 
source count rates in the 5--12\,keV and 1--5\,keV energy bands.
We only used colour information whenever the 1--20\,keV source count rate 
exceeded 3\,cts\,s$^{-1}$\,SSC$^{-1}$.

Note, that in the next Sections we use `intensity' as either the collimator 
corrected count rates ({\it HEAO}\,A-1 and {\it Ginga} ASM), or count rate 
per SSC ({\it RXTE} ASM).

Target of opportunity observations were performed at the {\it Australia 
Telescope Compact Array} ({\it ATCA}) during the outburst of 4U\,1630$-$47 in 
1996. The source was observed for 1.2, 3.1, 0.8 and 1.1\,hrs, on 
May 24, 25 , 26 and 27, respectively. All six antennae of the array were used,
with the longest baseline at 4.469\,km (array configuration 750D).
No polarisation data were taken. System gain and phase calibrations were 
carried out using the primary calibrators 1934$-$638 and 0823$-$500. The 
secondary calibrator 1718$-$469 was observed either immediately before or 
after 4U\,1630$-$47.

The radio data were analysed using the {\sc AIPS} software customised for the 
{\it ATCA} (Killeen 1993). Complete {\it uv} coverage was not 
possible for the source. Following correlation, the {\sc AIPS} task 
{\sc IMSTAT} was used to establish the 
background rms noise level from image pixels that were devoid of point 
sources. 

\section{Results}

\subsection{{\it HEAO}\,A-1 observations}

\begin{figure}
\centerline{\hbox{
\psfig{figure=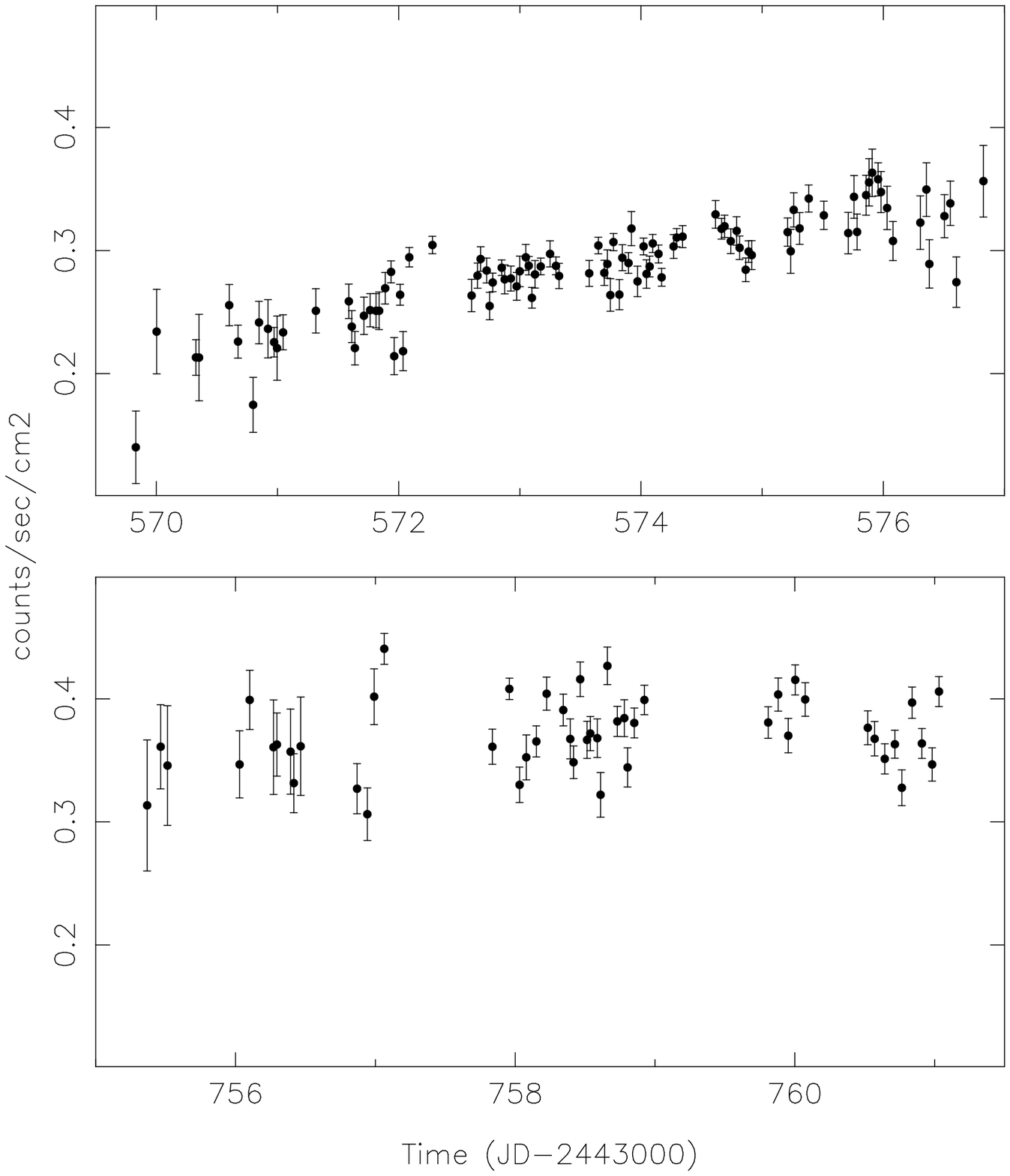,bbllx=65pt,bblly=156pt,bburx=534pt,bbury=739pt,angle=0,height=15cm}}
}
\caption{}
\label{heao_lc}
\end{figure}

The {\it HEAO}\,A-1 observations during 1978 March and September are shown
at the top and bottom of Fig.~\ref{heao_lc}, respectively. As can be seen in 
this figure, the source intensity increased from 
$\sim$0.2\,cts\,s$^{-1}$\,cm$^{-2}$ to $\sim$0.35\,cts\,s$^{-1}$\,cm$^{-2}$,
during the March observations. During the September observations the 
4U\,1630$-$47 intensity was approximately
constant at $\sim$0.37\,cts\,s$^{-1}$\,cm$^{-2}$, i.e.\ 
somewhat higher than during the March observations.
Source variability on time scales down to $\sim$30\,min are visible in
both light curves.

\subsection{Ginga ASM observations}

\begin{figure}
\centerline{\hbox{
\psfig{figure=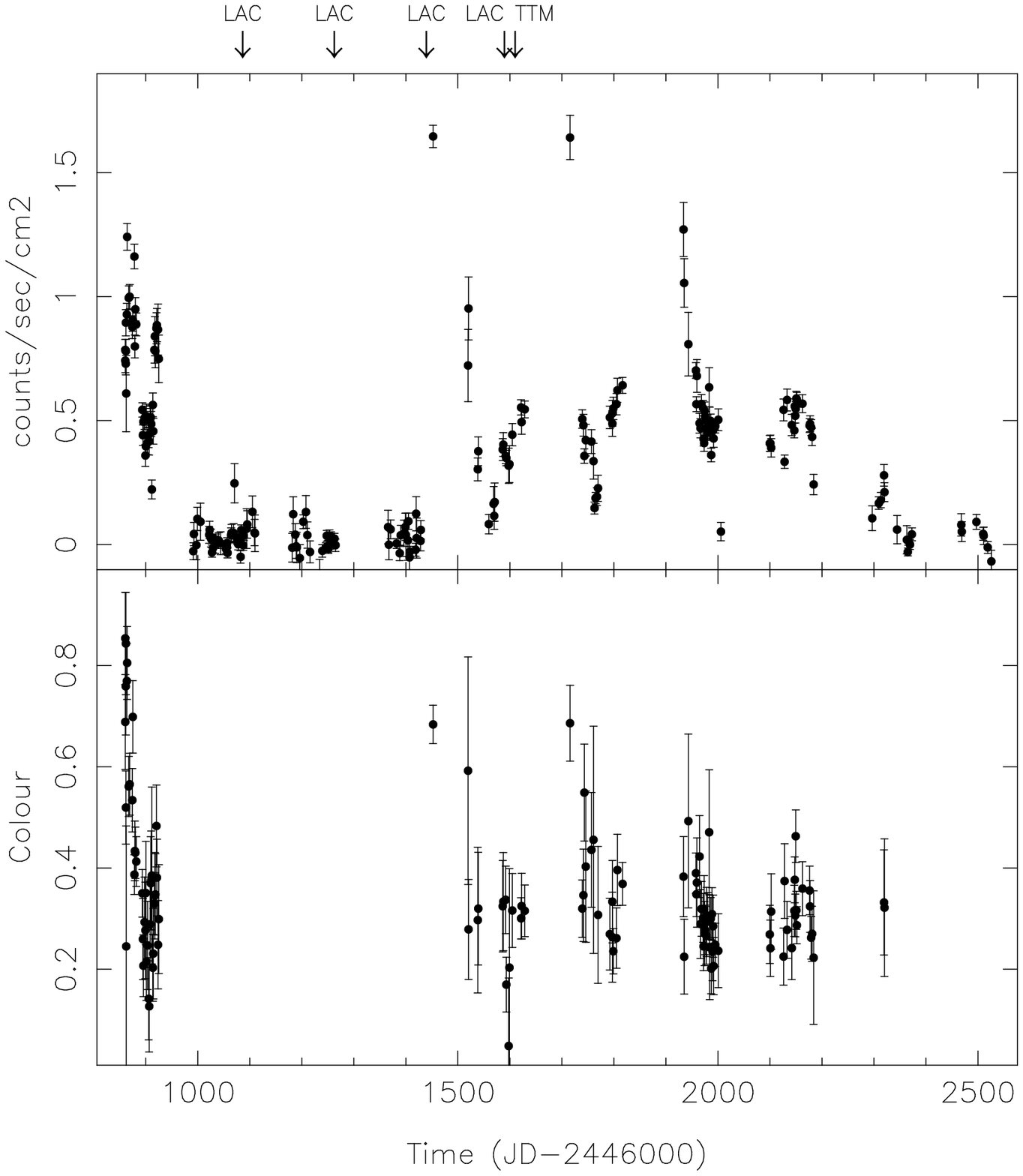,bbllx=65pt,bblly=194pt,bburx=534pt,bbury=739pt,angle=0,height=15cm}}
}
\caption{}
\label{ginga_lc}
\end{figure}

\begin{figure}
\centerline{\hbox{
\psfig{figure=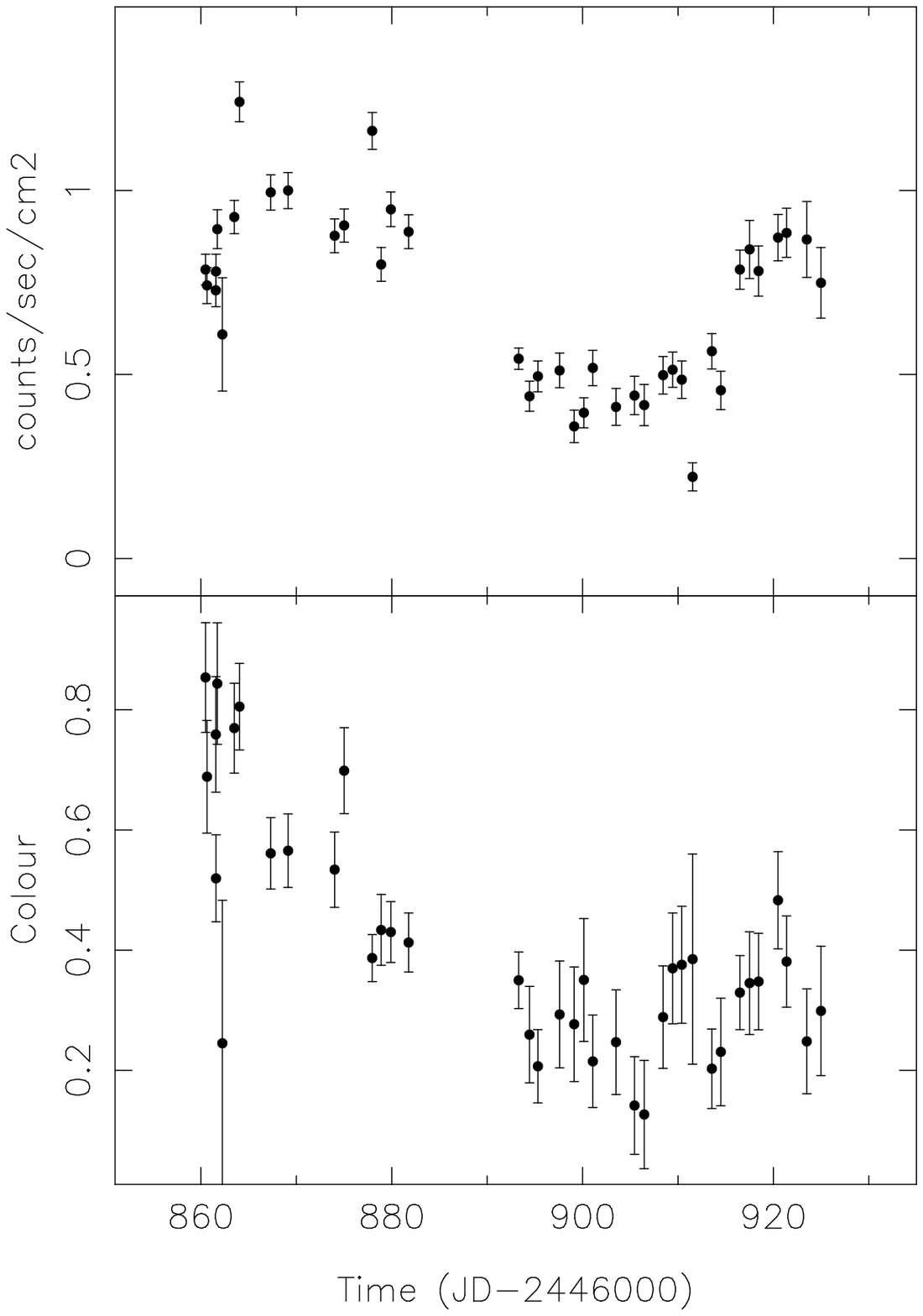,bbllx=65pt,bblly=194pt,bburx=534pt,bbury=739pt,height=15cm}}
}
\caption{}
\label{ginga_1987}
\end{figure}

The {\it Ginga} ASM light curve is shown in the upper panel of 
Fig.~\ref{ginga_lc}. At the start of the {\it Ginga} ASM observations, 
4U\,1630$-$47 was already in outburst, 
which had, therefore, started before 1987 March 5 (JD\,2\,446\,860). 
We refer to it as the 1987 outburst.
In Fig.~\ref{ginga_1987} we show the {\it Ginga} ASM light 
curve again, but now only for the 1987 outburst. 
The intensity is variable on time scales of a day during this outburst.
The intensity decreases erratically from around 1\,cts\,s$^{-1}$\,cm$^{-2}$ 
between March 5--26 to $\sim$0.5\,cts\,s$^{-1}$\,cm$^{-2}$ between 
April 17--28. After these dates the intensity was somewhat higher, at a level 
of $\sim$0.8\,cts\,s$^{-1}$\,cm$^{-2}$.
The lower limit on the outburst duration is 65~days.

The source was near or below the detection limits between 
1987 July 13 and 1988 September 24 (JD\,2\,446\,990--2\,447\,429).

From 1988 October 17 to 1991 March 4 (JD\,2\,447\,452--2\,448\,320), 
4U\,1630$-$47 showed previously
unreported extraordinary activity. Since this outburst started
in 1988, we refer to it as the 1988 outburst
(this outburst was previously denoted as the 1989 outburst, since the
outburst was thought to have started in 1989, see 
Parmar et al.\ [1997] and Sect.~4.1).
It showed two flares with intensities up to 
$\sim$1.5\,cts\,s$^{-1}$\,cm$^{-2}$ on 1988 October 17 (JD\,2\,447\,452) 
and 1989 July 7 (JD\,2\,447\,715), and probably a third one near 1990 February 10
(JD\,2\,447\,933). 
The overall light curve during the $\sim$870~days (i.e.\ $\sim$2.4\,yr) of 
activity is more or less symmetric and can be roughly described by a 
parabola. After each flare the flux decreased down to 
$\sim$0.1\,cts\,s$^{-1}$\,cm$^{-2}$, and then increased again 
within several days on the same time scale. This behaviour can be discerned 
during the whole activity period. Minima occurred roughly every $\sim$220~days, 
roughly similar to the separation of the flares.

The colour values during the 1987 and 1988 outburst are displayed in the
bottom panels of Figs.~\ref{ginga_lc} and \ref{ginga_1987}. 
As the 1987 outburst proceeded the colour values became lower, i.e.\ the
source spectrum became {\it softer}, until 40--50~days into the
{\it Ginga} ASM observations, when the colour value decrease flattened out.
During the observations when 4U\,1630$-$47 flared (1988 October 17 and 
1989 July 7), the colours were harder compared to the rest of the overall 
outburst. During the rest of the outburst the colour did not vary significantly,
but had values roughly between 0.2--0.5. 

No clear correlation between 1--6\,keV intensity and colour is discerned.
A clear correlation, however, exists between 
the 6--20\,keV intensity and colour (Fig.~\ref{ginga_hhid}): as the source
gets brighter in the 6--20\,keV energy range, the source spectrum hardens.

\subsection{{\it RXTE} ASM observations}

The top panel of Fig.~\ref{xte_lc} shows the {\it RXTE} ASM light curve.
The outburst started near 1996 March 11 (JD\,2\,450\,154) and rose to 
$\sim$19\,cts\,s$^{-1}$\,SSC$^{-1}$ within 
$\sim$15~days. We will denote this outburst as the 1996 outburst. 
The source slowly decayed erratically down to 
$\sim$13.5\,cts\,s$^{-1}$\,SSC$^{-1}$ on May 3 (JD\,2\,450\,207), and then 
started to increase again, slowly and erratically, up to
$\sim$22.5\,cts\,s$^{-1}$\,SSC$^{-1}$ on June 30 (JD\,2\,450\,265).
After this peak the source ended its outburst and quickly faded
in an exponential manner (with an e-folding time scale, $\tau_{\rm exp}$, of 
14.9$\pm$0.4~days) to within the detection limits in $\sim$70 days as shown in
Fig.~\ref{exp_dec_xte}. The overall light curve resembles an asymmetric 
`M'. The duration of the `plateau' phase was $\sim$100~days.

The colour variations during the 1996 outburst of 4U\,1630$-$47 are shown
in the bottom panel of Fig.~\ref{xte_lc}.
The source was soft at the start of the outburst, and became rapidly harder
(from $\sim$0.3 to $\sim$0.8) during the first half of the rise to maximum of 
the outburst. During the last part of the rise it remained more or less 
constant. As the outburst proceeded the overall colours increased,
although there is a large scatter in the colour values. This scatter is
mainly due to variability on a daily time scale in the 1.5--5\,keV intensity 
points. The overall colour values probably reached a maximum, at the end of 
the plateau phase of the outburst (i.e.\ at the peak intensity of the 
outburst). When the 1.5--12\,keV intensity started to decrease the 
spectrum did not show clear evidence for softening or hardening. There is,
however, an indication for a slight softening halfway through the decay of 
the outburst.

To study the rise to outburst in somewhat more detail, we plot in 
Fig.~\ref{start_1996} the intensities of the individual dwells of the 
three SSCs in three energy bands from February 2 to April 5
(JD\,2\,450\,115--2\,450\,179).
The dashed line in the three diagrams indicates the detection threshold of
the {\it RXTE} ASM. The outburst started more or less simultaneously in all
three energy bands, on March 11 (JD\,2\,450\,154). The intensity in the 
1.5--3\,keV band seemed to have reached its maximum by March 13 
(JD\,2\,450\,156). The 3--5\,keV intensity first rapidly increased up to 
March 13 and then continued to rise slowly to around March 20 
(JD\,2\,450\,163), after which it increased only slightly up to March 27
(JD\,2\,450\,170). The 5--12\,keV intensity increased more or less at the 
same rate until March 27.

In the hardness-intensity diagram the colour behaviour during the rise and
decay of the outburst results in a kind of anti-clockwise loop 
(Fig.~\ref{xte_hid}).
At the start of the outburst 4U\,1630$-$47 is in the lower left part of the
diagram. During the rise it moves to the middle right of the diagram. 
The source sits in this part of the diagram for most of the time, and
becomes (erratically) harder near the end of the outburst.
As the source decays, it moves to the left part of the diagram; at the
final decay stage the source moves to colours and intensity values similar
to those at the start of the outburst. 

\subsection{Radio observations}

During the 1996 outburst target of opportunity observations were performed 
with the {\it ATCA} from 1996 May 24 to 27 (JD\,2\,450\,227--2\,450\,230; 
indicated at the top of Fig.~\ref{xte_lc}).
No significant source was detected at the position of 4U\,1630$-$47. We 
derive 3$\sigma$ upper limits on the source flux of 4U\,1630$-$47 of 
5\,mJy at 1348\,MHz and 2\,mJy at 2354\,MHz.

\subsection{Outburst ephemeris}

The availability of the {\it Ginga} and {\it RXTE} ASM light curves enables us
to update the outburst ephemeris as presented in Parmar et al.\
(1997). Here we update the outburst times of the 
1987, 1988 and 1996 outbursts; for the other observed outburst times we 
use those reported in Parmar et al.\ (1997). 

The best estimate of the time of occurrence of the 1987 outburst is to take 
the peak intensity from the {\it Ginga} ASM data during this outburst.
This occurred near JD\,2\,446\,864. We assume a $\pm$50~day uncertainty, which
is consistent with the lower limit on the duration of the outburst.

For the 1988 outburst we take the time of the peak 
($\sim$1.6\,cts\,s$^{-1}$\,cm$^{-2}$) on JD\,2\,447\,452.
During one of the {\it Ginga} Large Area Counter (LAC) observations, 
$\sim$13\,days earlier (indicated in Fig.~\ref{ginga_lc}),
the source was reported to be near quiescence (Parmar et al.\ 1997). 
Note, that this implies that
a very fast rise to outburst must have occurred within this period.
We assign an uncertainty of $\pm$13~days to the time of the peak. We note 
that this is about 150 days earlier than estimated by
Parmar et al.\ (1997, see also Sect.~4.1)

For the 1996 outburst, it is difficult to give an estimate when the maximum 
occurred, since the XTE ASM light curve shows a plateau.
We therefore take the mid time of the outburst (JD\,2\,450\,220, i.e.\
1996 May 16) with an uncertainty equal to half the outburst duration 
(i.e.\ $\pm$60~days).

When we use these new values together with the other outburst times, 
a fit to a single outburst period is unacceptable 
($\chi^2_{\rm red}$ of 14, for 11 degrees of freedom [dof]) with a 
mean deviation of $\sim$80~days. 
If instead, we assume that the period and phase changed between the 
outburst which occurred in 1984 (Parmar et al.\ 1986) and the 
{\it Ginga} ASM observations, then the fits give more acceptable 
results ($\chi^2_{\rm red}$ of 4, for 9 dof). The period
before the change is 600.9$\pm$2.0~days, 
with $T_0 = {\rm JD}\,2\,440\,396$$\pm$9~days.
For these points the mean absolute deviation is $\sim$18~days. 
From just before the {\it Ginga} ASM observations it is 686.5$\pm$9.5~days with 
$T_0 = {\rm JD}\,2\,439\,222$$\pm$119~days. For these points the mean absolute 
deviation is $\sim$25~days.
In Fig.~\ref{arvind} we show the deviation of the outburst times 
from the expected times of the $\sim$601~day period before the period change, 
together with the results of the fit.

\section{Discussion}

\subsection{The 1977, 1987, 1988, and 1996 outbursts}

The 1977 outburst of 4U\,1630$-$47 started several months later than expected 
in 1977 November (Kaluzienski \&\ Holt 1977). Observations during 1978 
March (Share et al.\ 1978, Kaluzienski et al.\ 1978, this 
paper) and June (Sims \&\ Watson 1978) showed that 4U\,1630$-$47 was 
still active. We present additional observations with {\it HEAO}\,A-1 of 
4U\,1630$-$47 during this outburst obtained during 1978 September. 
The source intensity during the 1978 March and September {\it HEAO}\,A-1 
observations was an order of magnitude lower than at the peak near the start 
of the outburst (see Share et al.\ 1978, 
Kaluzienski et al.\ 1978).
We have shown that the source intensity was somewhat higher during the 
September with respect to the intensity measured during the 1978 March 
observations. This indicates, that either 4U\,1630$-$47 had started a new 
outburst (which is inconsistent with the $\sim$600~day outburst recurrence
interval), or that 4U\,1630$-$47 was still active $\sim$10~months after the 
outburst had started.
As we have shown, 4U\,1630$-$47 was active for at least $\sim$2.4~years
after the start of the 1988 outburst (see also below); we therefore favor 
the latter possibility, i.e.\ that 4U\,1630$-$47 was active for at least 
$\sim$10~months during the outburst which started in 1977.

Observations with the {\it Ginga} ASM started when 4U\,1630$-$47 was already 
active in 1987. This outburst must have started, therefore, before 1987 
March 5. Note that this is more than three months earlier than the outburst
peak time estimated by Parmar et al.\ (1986),
who used one single {\it Ginga} LAC observation.
They noted that the source spectrum during this observation (on 1987 July 13)  
resembled the spectrum taken at the end of the 1984 outburst 
(see Parmar et al.\ 1986), and therefore concluded that 
during the {\it Ginga} LAC observations 4U\,1630$-$47 was near the 
end of the outburst.
From the {\it Ginga} ASM observations we infer that the outburst duration is 
at least 65~days. If one takes into account that the {\it Ginga} LAC 
observation on 1987 July 13 was made near the end of the outburst, the 
outburst duration is at least $\sim$120~days. During the outburst the 
spectrum {\it softened}. 

New outburst behaviour is displayed during the 1988--1991 {\it Ginga} ASM 
observations of 4U\,1630$-$47. Its outburst activity lasted for up to
$\sim$2.4 years\footnote{It is interesting to note that, recently 
Bloser et al.\ (1996) reported a $\sim$3.5$\sigma$ detection of a transient
hard X-ray (20--100\,keV) event with BATSE in the direction of 
4U\,1630$-$47 during 1991 May (JD\,2\,448\,380--2\,448\,410). This indicates 
that the outburst activity may have lasted even longer than inferred from 
the {\it Ginga} ASM observations.}, i.e.\ even longer than its outburst 
recurrence time.
It showed a parabolic shaped outburst with large flares and quasi-periodic 
increases and decreases in intensity by a factor $\sim$5 on a time scale
of $\sim$220~days. During the flares the spectrum hardens compared to the 
rest of the outburst. The first flare, observed at 1988 October 17, 
occurred 13 days after a non-detection 
(L$_{\rm X}$$\la$10$^{36}$\,erg\,s$^{-1}$) with the {\it Ginga} LAC
(see Parmar et al.\ 1997). This shows that 4U\,1630$-$47
rose to a peak brightness, i.e.\ more than two orders of magnitude,
within $\sim$13~days. We infer, therefore, that the 1988 outburst 
started sometime between the {\it Ginga} LAC and {\it Ginga} ASM flare 
observations.

Observations of 4U\,1630$-$47 with the {\it Ginga} LAC on 1989 March 2--4 
(Parmar et al.\ 1997) and the COMIS/TTM instrument onboard 
{\it Mir/Kvant} (in 't Zand 1992) on 1989 March 24--25, i.e.\ 
$\sim$22~days later (indicated in Fig.~\ref{ginga_lc}), showed 
that the flux increased
between these two observations. This is confirmed by our {\it Ginga} ASM
observations, which show a systematic increase with a factor of $\sim$6
from 1989 February 2 to April 11. From the two observations 
Parmar et al.\ (1997) inferred that the TTM observation
was obtained during the rise of the outburst, and that the peak of the outburst
occurred about halfway between the two observations. However, as shown
above, the actual peak of the outburst occurred $\sim$5~months earlier.
Similar behaviour can be seen during the 1971 outburst\footnote{We note that 
the 1971 outburst had already {\it started} at the end of 1970, see figure 1b 
in Priedhorsky (1986).} (Forman, Jones, \&\ Tananbaum 1976, 
Jones et al.\ 1976, Priedhorsky 1986); after reaching its peak, 
the source intensity decreased to almost the pre-outburst intensity, but 
increased again to about 80\%\ of its peak flux.
We note, therefore, that assigning outburst times to incompletely covered 
outbursts of 4U\,1630$-$47 should be treated with some caution. 

{\it Ginga} ASM observations between 1987 July 13 and 1988 September 24 
indicate the source was near or below the detection limit. Indeed, 
{\it Ginga} LAC observations on 
1987 October 18 and 1988 April 11 (Parmar et al.\ 1997; also
indicated in Fig.~\ref{ginga_lc}) show the 
source at a luminosity of $\sim$10$^{36}$\,erg\,s$^{-1}$ (at 10\,kpc), i.e.\ 
more than 2 order of magnitude below the observed peak outburst luminosity. 
Since the `true' quiescent luminosity is probably lower
than $\sim$10$^{33}$\,erg\,s$^{-1}$ (Parmar et al.\ 1997), 
the {\it Ginga} LAC observations between the 1987 and 1988 outburst indicate 
that there was still considerable activity.

In 1996, a new outburst of 4U\,1630$-$47 was discovered by {\it RXTE} 
(Marshall 1996; Levine et al.\ 1996b). 
The outburst starts simultaneously in the three {\it RXTE} ASM energy bands.
The light curve shows a fast rise ($\la$15~days), a
plateau phase for $\sim$100~days (although the plateau was not 
completely flat), and an exponential decay with an e-folding time scale of
$\sim$15~days. The decay timescales of SXT outbursts are distributed generally 
between $\sim$5--100~days, peaking between $\sim$20--30~days (see Chen 
et al.\ 1996). So, the decay of the 1996 outburst may be regarded as a fast 
one.

The colours of 4U\,1630$-$47 show an interesting behaviour 
during the rise and decay of the outburst, performing an anti-clockwise 
loop in the hardness-intensity diagram. At the start of the
outburst 4U\,1630$-$47 shows a soft spectrum, which rapidly becomes harder
during the first part of the rise to maximum, and then flattens during the
last part of the rise to maximum. As the intensity started to decay, the 
colours did not change significantly; only halfway through the decay the colours 
decreased slightly, comparable to the colours near the beginning of the
plateau phase. This behaviour is due to the following:
during the first days of the rise the soft intensity (1.5--5\,keV) rapidly 
increases, while the hard intensity (5--12\,keV) rises more slowly; this
gives the source a soft spectrum. Subsequently the soft intensity
rises slower (or remains constant in the 1.5--3\,keV band), while
the hard intensity increases at the same rate as before. This results in
the spectrum becoming rapidly harder. 
Such colour behaviour during the start of an outburst might be expected 
when an instability in the disk starts in the outer or middle regions
of the disk, and spreads inwards (so-called outside-inside outburst).
Such models for BH SXTs have been discussed by, e.g.\ Lasota et al.\ 
(1996) in which the inner advection dominated region 
(see Lasota et al.\ 1996, Narayan 1997, and references 
therein) collapses, so that one would expect soft X-rays to reach their 
maximum earlier than the hard X-rays (assuming that the formation region
for the hard X-rays is generally closer to the black hole than
that for the soft X-rays).

During the plateau phase the spectrum of 4U\,1630$-$47 generally became 
harder, although the scatter increases towards the end of the plateau, due to
variations in the 1.5--5\,keV intensity. This hardening towards the end of
the plateau phase is consistent with the report of a detection of 
4U\,1630$-$47 with the Burst and Transient Source Experiment (BATSE), by 
earth occultation measurements in June 1996 
(Shrader 1996). However, the detection was only at the 3.5--4.5$\sigma$ 
level, i.e.\ $\sim$100\,mCrab, and there may
be source confusion with GRO\,J1655$-$40
(Shrader 1996, private communication), which was in outburst 
(e.g.\ Remillard et al.\ 1996) at a level close to 1 Crab during that 
time (Shrader 1996).
The {\it hardening} during the 1996 outburst is opposite to that seen during
the 1987 outburst.

Plateau X-ray lightcurves like the 1987 and 1996 outbursts of 4U\,1630$-$47
have been seen previously in a wide variety of LMXBs, that contain both 
neutron stars and black holes; such a shape can be seen in different energy 
bands (Chen et al.\ 1997). Their durations are on the order of 
$\sim$10--300~days. Other outburst shapes are those with fast rise and
exponential decay (so-called FREDs) or are more erratic (see Chen et al.\ 
1997). We note that some outburst light curves of 4U\,1630$-$47 
(Priedhorsky 1986) resembles a FRED light curve. 

\subsection{Outburst time scales}

The durations of the 1987 and 1996 outbursts are comparable to the
durations of three previous outbursts (1971, 1972, 1984, i.e.\ between 
$\sim$95--120~days (Chen et al.\ 1997, see also Forman, Jones \&\ 
Tanabaum 1976, Jones et al.\ 1976, Priedhorsky 1986). 
Two other previous outbursts (1974 and 1976) had a duration of 
$\sim$240--250~days (Chen et al.\ 1997, see also Jones et al.\ 
1976, Priedhorsky 1986). So, outbursts with durations of several 
months seem to occur normally every outburst cycle in 4U\,1630$-$47,
whereas activity lasting for up to $\sim$2.4~year is rare. 

Bi-modal outburst activity is also known in the cataclysmic variable class,
the SU\,UMa stars, which show apart from frequent `normal' outburst
(lasting several days), less frequent `superoutbursts' (lasting several
weeks). It has already been shown that various analogies exist between 
SXTs and SU\,UMa stars (Lasota 1996, \mbox{Kuulkers}, Howell \&\ Van Paradijs
1996). On the other hand, outbursts with different durations have
also been seen in peculiar Be/X-ray binaries such as V0332+53 (weeks 
vs.\ several months, see Whitlock 1989, Makishima et al.\ 1990)
and A0538$-$66 (durations between $\sim$0.1--14~days, Skinner et al.\ 
1980). 
Since the optical counterpart of 4U\,1630$-$47 is undetected, the mass ratio 
of 4U\,1630$-$47 is unknown. It will be interesting to see whether it 
is as extreme as in certain other SXTs and SU\,UMa stars 
(see \mbox{Kuulkers} et al.\ 1996, and
references therein), and/or if the secondary star is of Be-type
(see also Sect.~4.4), or if it is a system unrelated to
these binaries. We note that recently 
GRS\,1915+105, a system showing similar behaviour as 4U\,1630$-$47 
(see discussion below) was proposed to be a high-mass X-ray binary 
based on IR measurements (Mirabel et al.\ 1997).

While most of the outbursts of 4U\,1630$-$47 have durations 
of $\sim$95--250~days, we show that during the 1988 outburst the X-ray 
intensity showed flares and quasi-periodic intensity variations on time scales 
of about $\sim$220~days.
Such a 100--200~days time scale is also observed in some other BHXTs.
Glitches, or secondary maxima, often occur $\sim$100 days after the peak
of the outburst in FRED light curves (see Chen, Livio \&\ Gehrels 
1993, Augusteijn, \mbox{Kuulkers} \&\ Shaham 1993, Chen et al.\ 
1997, and references therein). Optical reflares have been seen
to occur after the main outburst on a similar timescale (Callanan et al.\
1995, Chevalier \&\ Ilovaisky 1995, Bailyn \&\ Orosz 1995). 

GRS\,1915+105 and GRO\,J1655$-$40 are galactic X-ray sources which have 
shown jets moving at relativistic 
(apparently superluminal) speeds during radio outbursts 
(Mirabel \&\ Rodr\'{\i}guez 1994,
Harmon et al.\ 1995, Hjellming \&\ Rupen 1995,
Tingay et al.\ 1995); GRS\,1716$-$249 may be a similar type of 
source (Hjellming et al.\ 1996). 
GRO\,J1655$-$40 has been shown dynamically to contain a black hole
(Bailyn et al.\ 1995).
In GRO\,J1655$-$40 several outburst peaks every 
$\sim$120~days were reported during {\it GRO/BATSE} observations in 1994 
(Harmon et al.\ 1995).
Similarly, every $\sim$100~days GRS\,1716$-$249 (see Hjellming et al.\ 
1996) and GRS\,1915+105 (see Foster et al.\ 1996) also show 
outburst peaks in their light curves. 
We therefore conclude that a 100-200~days timescale is evident in a
number of sources.
 
It is interesting to note, that the superluminal sources have been also 
seen to show activity at similar time scales as the outburst recurrence times
of 4U\,1630$-$47.
Three intense outbursts have been observed from GRS\,1915+105 with 
{\it GRO/BATSE} (see Harmon et al.\ 1995). 
These outbursts started in 1992 August (Castro-Tirado et al.\ 1992),
1994 March (Sazonov, Sunyaev \&\ Lapshov 1994), and 1995 October 
(Harmon et al.\ 1995), which may suggest a recurrence interval of 
$\sim$19~months, i.e.\ close to the outburst recurrence interval of 
4U\,1630$-$47. We note that the start of the next intense outburst from 
GRS\,1915+105 is expected in 1997 May. 
A similar time scale has been noted between the occurrence of 
radio flares in GRO\,1655$-$40 and GRS\,1716$-$24, i.e.\ $\sim$16~months
and a couple of years, respectively (Hjellming et al.\ 1996). 

\subsection{Outburst activity}

We have shown that 4U\,1630$-$47 remained active for up to 2.4~years 
during the 1988 outburst and exhibited large variations on timescales of a 
day to months, similar to these three sources. Moreover, observations
in between the 1987 and 1988 outburst (Parmar et al.\ 1997) suggest
that `true' quiescence had not been reached, which points to inter-outburst 
activity. Lower level inter-outburst activity, i.e.\ at intensities $\sim$2 
order of magnitude lower than during the observations in between the 
1987 and 1988 outburst, was also reported by Parmar et al.\ (1997).

The X-ray light curves in different energy bands 
of GRS\,1915+105 (Paciesas et al.\ 1995, Foster et al.\ 1996, 
Greiner, Morgan \&\ Remillard 1996, Sazonov et al.\ 1996)
GRO\,J1655$-$40 (Sazonov et al.\ 1996, Tavani et al.\ 1996), 
and GRS\,1716$-$249 (Hjellming et al.\ 1996)
show considerable outburst and inter-outburst variability on different 
timescales. The intensity in these sources varies from day to day, and their 
outbursts may last up to several months to years. Again, these three sources
show behaviour similar to 4U\,1630$-$47.

\mbox{Kuulkers} et al.\ (1997) have recently shown that at times
4U\,1630$-$47 exhibits similar fast X-ray timing 
and X-ray spectral behaviour to that seen in GRS\,1915+105 and GRO\,J1655$-$44
(Ebisawa 1996, Zhang et al.\ 1997).
Since, at certain times, the X-ray timing and spectra and the 
long-term activity resemble those of the Galactic superluminal 
sources, we again speculate, similar to \mbox{Kuulkers} et al.\ (1997), that 
4U\,1630$-$47 may exhibit considerable radio activity with associated 
relativistic radio jets during outburst.

Radio observations were performed when the 1996 outburst was half way through.
We found, however, no significant radio source at the position of
4U\,1630$-$47. This means that either 4U\,1630$-$47 was not active at radio
frequencies throughout the outburst, or was active at other times.
We note that there is no clear correlation between the 
hard X-ray (20--100\,keV) outburst and radio activity of GRS\,1915+105 and 
GRO\,J1655$-$40 (see Foster et al.\ 1996, Tavani et al.\ 1996), i.e.\ not
every soft/hard X-ray outburst is accompanied by a radio outburst.
So, future radio monitoring during an X-ray outburst is required to see 
whether 4U\,1630$-$47 exhibits radio activity similar to the 
superluminal sources.  

\subsection{Outburst recurrence times}

We find a phase and period change in the outburst recurrence of 4U\,1630$-$47.
The period may have changed from $\sim$600~days to $\sim$690~days between the 
outbursts which occurred in 1984 and 1987. 
Lasota, Narayan, \&\ Yi (1996) recently noted
that $\sim$600~days might be the {\it orbital} period of 4U\,1630$-$47. Since 
orbital period changes of $\sim$90~days are unlikely, 
the outburst recurrence interval is not directly related to orbital variations
(see also Kitamoto et al.\ 1993).
If the period of $\sim$690~days remains stable, we infer that the next 
outburst will start near JD\,2\,450\,845 (1998 January 31).

We note (see also 
Amnuel, Guseinov \&\ Rakhamimov 1979) that 4U\,1630$-$47 was
detected during a rocket flight in 1969 (Cruddace et al.\ 1972).
The observation (4U\,1630$-$47 is designated as GX\,337+0) on June 14 
is indeed near the observed outburst time of the 1969 outburst
(see Parmar et al.\ 1995, 1997). During an earlier rocket 
flight in 1965 (Friedman, Byram \&\ Chubb 1967) a source (labelled
Nor\,XR-1) was detected near the position 
(see Cruddace et al.\ 1972) of GX\,337+0/4U\,1630$-$47. If
Nor\,XR-1 and GX\,337+0/4U\,1630$-$47 are the same source then
the time of the detection (April 25) is inconsistent with the expected
outburst times as derived from our outburst ephemeris, i.e.\
JD\,2\,438\,593 (1964 July 16) or JD\,2\,439\,194 (1966 March 9).
This means that either 4U\,1630$-$47 was still active 
after the outburst which started in 1964 (similar to the 1977 and 1988 
outbursts), or it had a different recurrence time scale and/or phase with 
respect to the 1969--1984 epoch, or the source was not GX\,337+0/4U\,1630$-$47.

Evidence for unstable outburst recurrence intervals has
been previously reported for the SXT Aql\,X-1 (known to contain a neutron star) 
by Priedhorsky \&\ Terrell (1984); change from $\sim$123 to $\sim$127 
days) and Kitamoto et al.\ (1993); $\sim$309~days vs.\ $\sim$125
days). Priedhorsky (1986) suggested that such period (and probably 
phase changes) may be related to a similar phenomenon in SU\,UMa 
stars. SU\,UMa stars show, apart from 
short normal outbursts, longer superoutbursts which occur quasi-periodically 
for 10--20 cycles with an rms of 5--10\%\/; occasionally the recurrence 
time changes to another value (Vogt 1980). 

Deviations in the recurrence times, and also in the intensities at the 
peak of the outburst and its duration (see also Sect.~4.2)
have also been noted on the Be/X-ray binary A0538$-$66 
(A0535$-$668; Skinner et al.\ 1980). 
Some outbursts occurred $\sim$1--2~days later than 
expected. The deviations were either attributed to phase jitter
or to a change in period occurring around the time of an exceptionally 
long outburst. The outburst times of 4U\,1630$-$47 since 1987 are 
inconsistent with only 
phase jitter, unless the phase is constantly changing.
A0538$-$66 also undergoes transitions between high and low
activity on time scales of about a year (Skinner et al.\ 1980,
Pakull \&\ Parmar 1981). Moreover, interoutburst (optical) activity 
has been found in this system as well (Densham et al.\ 1983). 
We have argued in this paper that 4U\,1630$-$47 shows times of considerable
inter outburst activity (at L$_{\rm X}$$\sim$10$^{36}$\,erg\,s$^{-1}$),
whereas at other times the source is very quiet 
($\la$10$^{33}$\,erg\,s$^{-1}$, Parmar et al.\ 1997), and is
therefore reminiscent of the behaviour of A0538$-$66 described above.
However, the outbursts of A0538$-$66 are believed to recur every
orbital period, whereas we have shown that the outburst recurrence time of 
4U\,1630$-$47 cannot be related to orbital variations, if the recurrence 
interval did indeed change by $\sim$90~days between the 1984 and 1987 outburst.
As noted in Sect.~4.2, optical identification of 4U\,1630$-$47 would 
therefore shed more light on establishing the nature of the system. 

\subsection{Outburst activity}

Why does 4U\,1630$-$47 shows periods of recurring relatively short
outbursts and periods of long-term activity? A possible explanation
is variations in the average mass transfer rate from the secondary. 
Recently, Orosz \&\ Bailyn (1997)
inferred that GRO\,J1655$-$40 is probably close to being a persistent source.
Systems accreting far below the critical rate show less
frequent outbursts than systems accreting close to the 
critical rate, such as GRO\,1655$-$40. So, when mass transfer from 
the secondary in 4U\,1630$-$47 is very close to its critical rate, it might
show long-term activity as seen during the 1977 and 1988 outbursts.
When the average transfer rate is lower, the outbursts are shorter and
`true' quiescence may perhaps be reached in between outbursts.

On the contrary, 
Belloni et al.\ (1997) inferred from {\it RXTE} observations 
that the average mass accretion rate in GRS\,1915+105 is above the critical 
rate, and that outbursts in this system are caused by drops in the accretion 
rate, so that the accretion disk can become unstable. This could be the 
explanation for the long-term activity seen in 4U\,1630$-$47, but it is,
however, hard to reconcile with the `normal' outburst behaviour of
4U\,1630$-$47. 

Another mechanism for rapid short repeating outbursts (or long-term 
activity) may be irradiation induced mass transfer from the secondary.
An example are the three outbursts of GRO\,J1655$-$40 which occurred 
shortly after each other in 1994. It was argued by Wu (1997) that 
the first two outbursts of GRO\,J1655$-$40 may have been triggered by disk 
instabilities (or a disk instability triggered by enhanced mass transfer,
Lasota et al.\ 1996), 
but that the even stronger third outburst had to be triggered by
another mechanism, since most of the material would have been depleted 
by the two outbursts. Wu (1997) proposed an irradiation induced mass 
transfer instability of the secondary star due to the energetic preceding 
outbursts as a plausible model. We note that irradiation induced 
mass transfer increases have also been invoked to explain the secondary maxima 
or so-called glitches during FRED light curves of some BH SXTs 
(e.g.\ Chen et al.\ 1993, Augusteijn et al.\ 1993).
We suggest that such a model might also explain the flares and possibly the 
quasi-periodic intensity behaviour during the long 1988 outburst of 
4U\,1630$-$47. 

\section{Conclusions}

\begin{description}
\item For the first time we present detailed 
outburst light curves and colour behaviour of the black-hole candidate
4U\,1630$-$47. The source shows different and complex outburst behaviour during 
the different outbursts we have studied, i.e.\ the outbursts which started in 
1977, 1987, 1988 and 1996.
\item 4U\,1630$-$47 showed previously unreported long-term outburst 
activity of up to $\sim$2.4~years after the start of its outbursts in 1977 and 
1988. The 1988 outburst showed several large flares on order of a few days,
and quasi-periodic intensity variations on time scales of order 220~days.
The long term activity resembles that seen in GRO\,J1655$-$40, GRS\,1915+105,
and GRS\,1716$-$24.
\item During the outbursts in 1987 and 1996 4U\,1630$-$47 showed 
different colour behaviour during each outburst. During the outburst in 1987 
the X-ray spectrum softened, whereas during the one in 1996 it hardened,
as the outbursts proceeded. 
\item 4U\,1630$-$47 showed a phase and period change in its 
outburst recurrence interval between the outbursts which occurred in 1984 and
1987. Before 1987 4U\,1630$-$47 showed outbursts every $\sim$600~days, 
whereas after it the outbursts recurred with an interval of $\sim$690~days.
Such a large interval change implies that the recurrent outburst
behaviour cannot be directly related to the orbital period. If the next outburst
occurs $\sim$690~days after the 1996 event, we predict
that this will occur near 1998 January 31.
\end{description}

\section*{acknowledgements}

EK thanks Phil Charles for useful discussions.
This research has made use of {\it RXTE} ASM data obtained through 
the High Energy Astrophysics Science Archive Research Center Online Service, 
provided by the NASA/Goddard Space Flight Center. The Australian Telescope
is funded by the Commonwealth of Australia for operation as a National
Facility managed by CSIRO.

\newpage

\begin{center}
{\bf Figure Captions}
\end{center}

{\bf Figure 1.} {\it HEAO} A-1 0.25--25\,keV light curves during 1978 
March 1--8 (top) and September 2--8 (bottom).\\

{\bf Figure 2.} {\it Ginga} ASM 1--20\,keV light curve (top) and colour 
(6--20\,keV/1--6\,keV) curve (bottom) from 1987 March 5 (JD\,2\,446\,861) to 
1991 September 26 (JD\,2\,448\,526). Note the previously unreported long-term 
X-ray activity between 1988 October ($\sim$JD\,2\,447\,452) and 1991 March
($\sim$JD\,2\,448\,320). The times of the {\it Ginga} LAC (Parmar et al.\ 1997)
and the {\it Mir/Kvant} TTM observations (in 't Zand 1992) are indicated.\\

{\bf Figure 3.} {\it Ginga} ASM light (top) and colour (bottom) curve during the 
1987 outburst from March 5 (JD\,2\,446\,861) to May 8 (JD\,2\,446\,924).\\

{\bf Figure 4.} Hardness (or colour) vs.\ 6--20\,keV intensity for all 
{\it Ginga} ASM observations, whenever the intensity exceeded 
0.2\,cts\,s$^{-1}$\,cm$^{-2}$. Note the correlation between hardness 
and intensity: when the intensity increases the source spectrum gets harder.\\

{\bf Figure 5.} {\it RXTE} ASM 1--12\,keV light curve (top) and colour 
(5--12\,keV/1--5\,keV) curve (bottom) from 1996 January 5 (JD\,2\,450\,088) 
to 1996 November 7 (JD\,2\,450\,395).
The times of the target of opportunity observations with the {\it Australian 
Telescope Compact Array} ({\it ATCA}) are indicated.\\

{\bf Figure 6.} {\it RXTE} ASM light curve in three energy bands, i.e.\ 1.5--3\,keV (top),
3--5\,keV (middle) and 5--12\,keV (bottom), near the start of the 1996 
outburst. Data from individual SSCs are plotted. The start of the outburst
is approximately the same in all three energy bands, i.e.\ near 
1996 March 11 (JD\,2\,450\,154).\\

{\bf Figure 7.} Hardness (or colour) vs.\ 1--12\,keV intensity for 
the {\it RXTE} ASM observations, whenever the intensity exceeded 
3\,cts\,s$^{-1}$\,SSC$^{-1}$. The evolution of the data points
during the outburst (starting from the lower left corner of the diagram) is
indicated schematically by arrows.\\

{\bf Figure 8.} {\it RXTE} ASM 1--12\,keV light curve of the end of the 1996
outburst. The drawn line shows the exponential fit to the data
with an e-folding time scale of 14.9~days.\\

{\bf Figure 9.} The time residual in days of all outburst times after a linear fit 
to the first eight outburst times to cycle number, vs.\ cycle number. 
At the top we give the corresponding time in years. The error bars reflect
the uncertainties in deriving the outburst start times (see text and 
Parmar et al.\ 1997). The dashed line shows the linear fit to the
last five residual outburst times to cycle number.
The recurrence interval has changed its period and phase between the
1984 (cycle 9) and 1987 outbursts (cycle 11).

\bsp % ``This paper has been produced using the ...''

\label{lastpage}

\newpage

\begin{figure}
\centerline{\hbox{
\psfig{figure=kuulkers_1630_fig4.ps,bbllx=66pt,bblly=194pt,bburx=534pt,bbury=702pt,angle=0,height=15cm}}
}
\caption{}
\label{ginga_hhid}
\end{figure}

\begin{figure}
\centerline{\hbox{
\psfig{figure=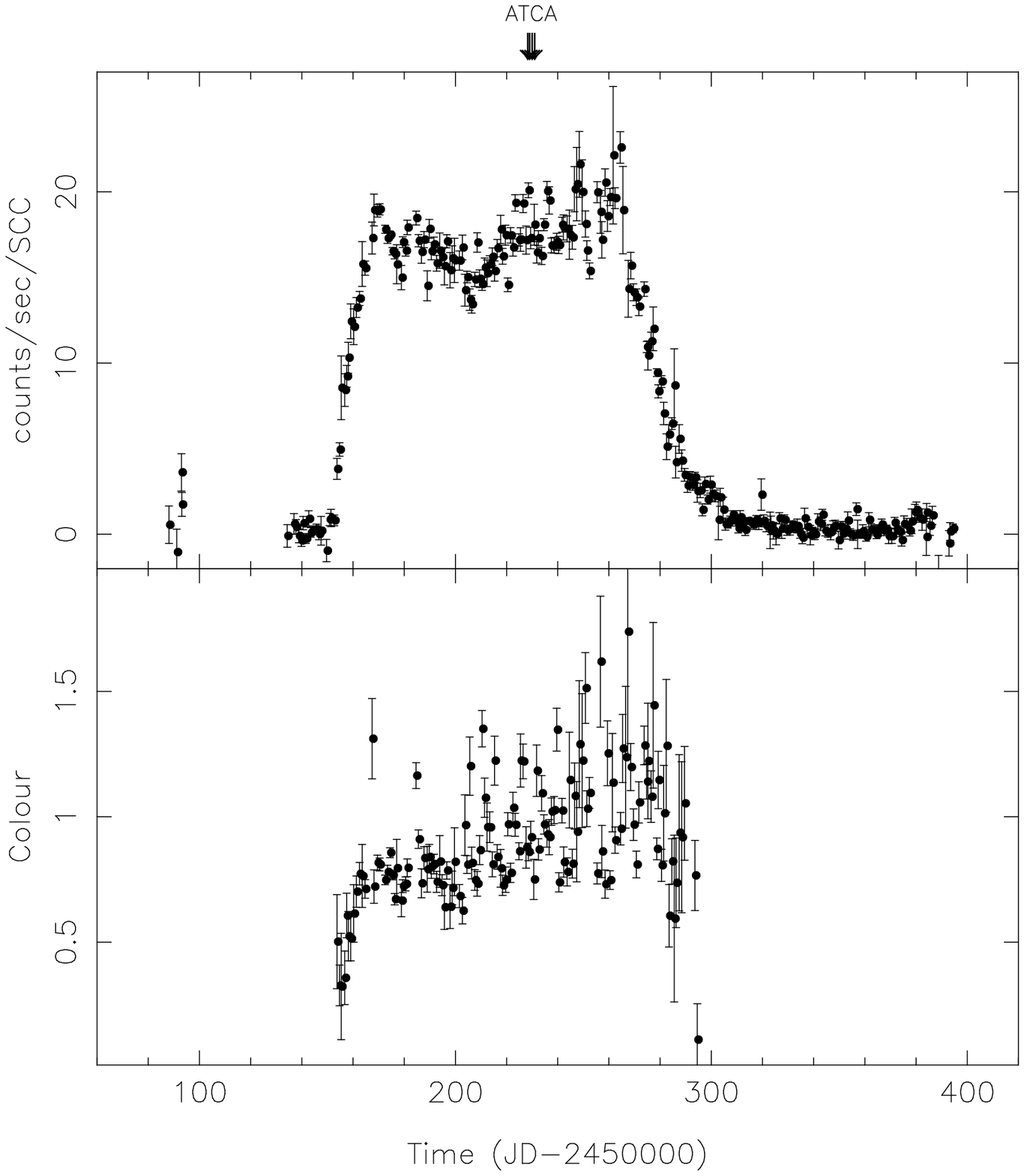,bbllx=65pt,bblly=194pt,bburx=534pt,bbury=739pt,angle=0,height=15cm}}
}
\caption{}
\label{xte_lc}
\end{figure}

\begin{figure}
\centerline{\hbox{
\psfig{figure=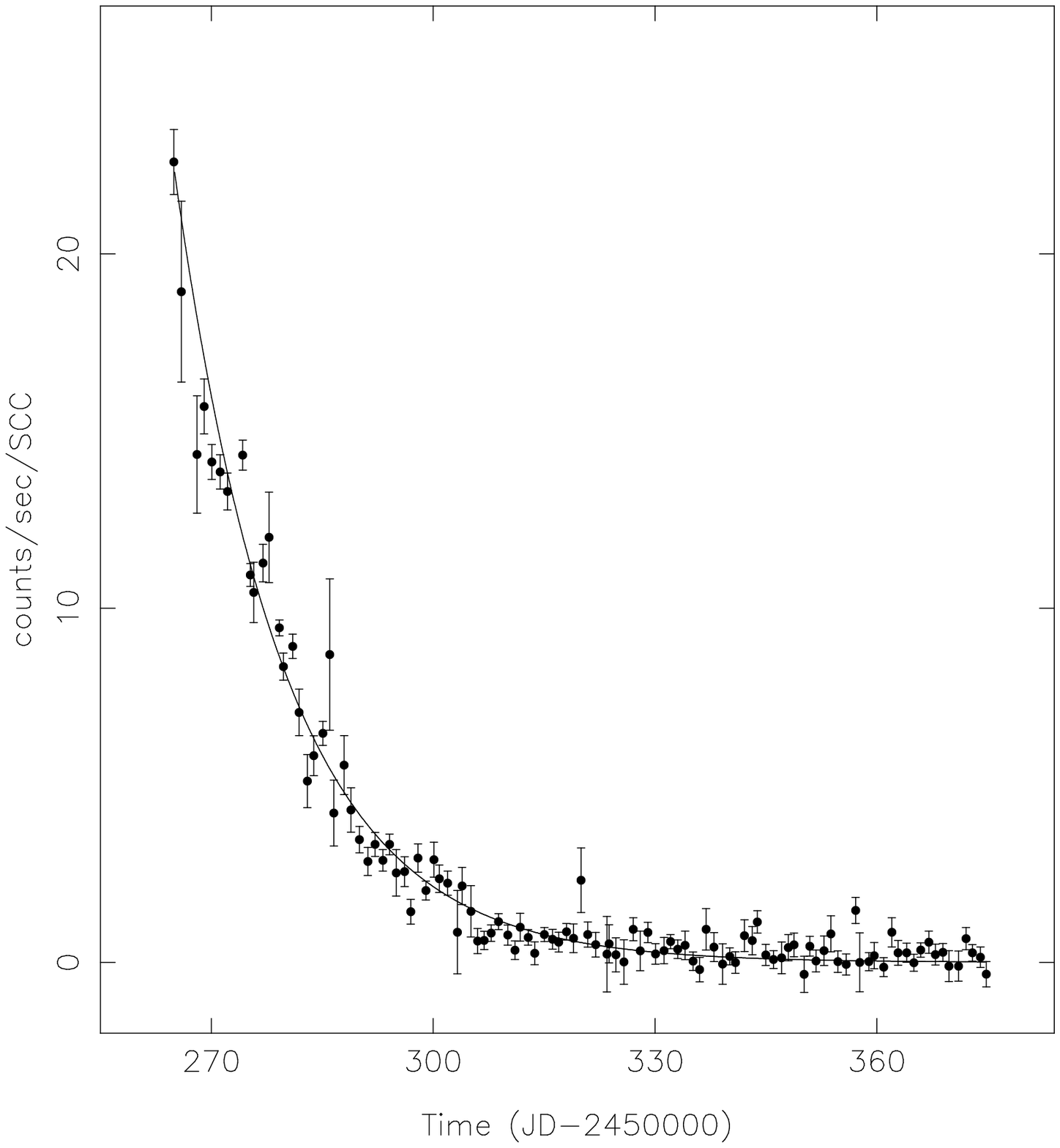,bbllx=65pt,bblly=194pt,bburx=534pt,bbury=739pt,height=15cm}}
}
\caption{}
\label{exp_dec_xte}
\end{figure}

\begin{figure}
\centerline{\hbox{
\psfig{figure=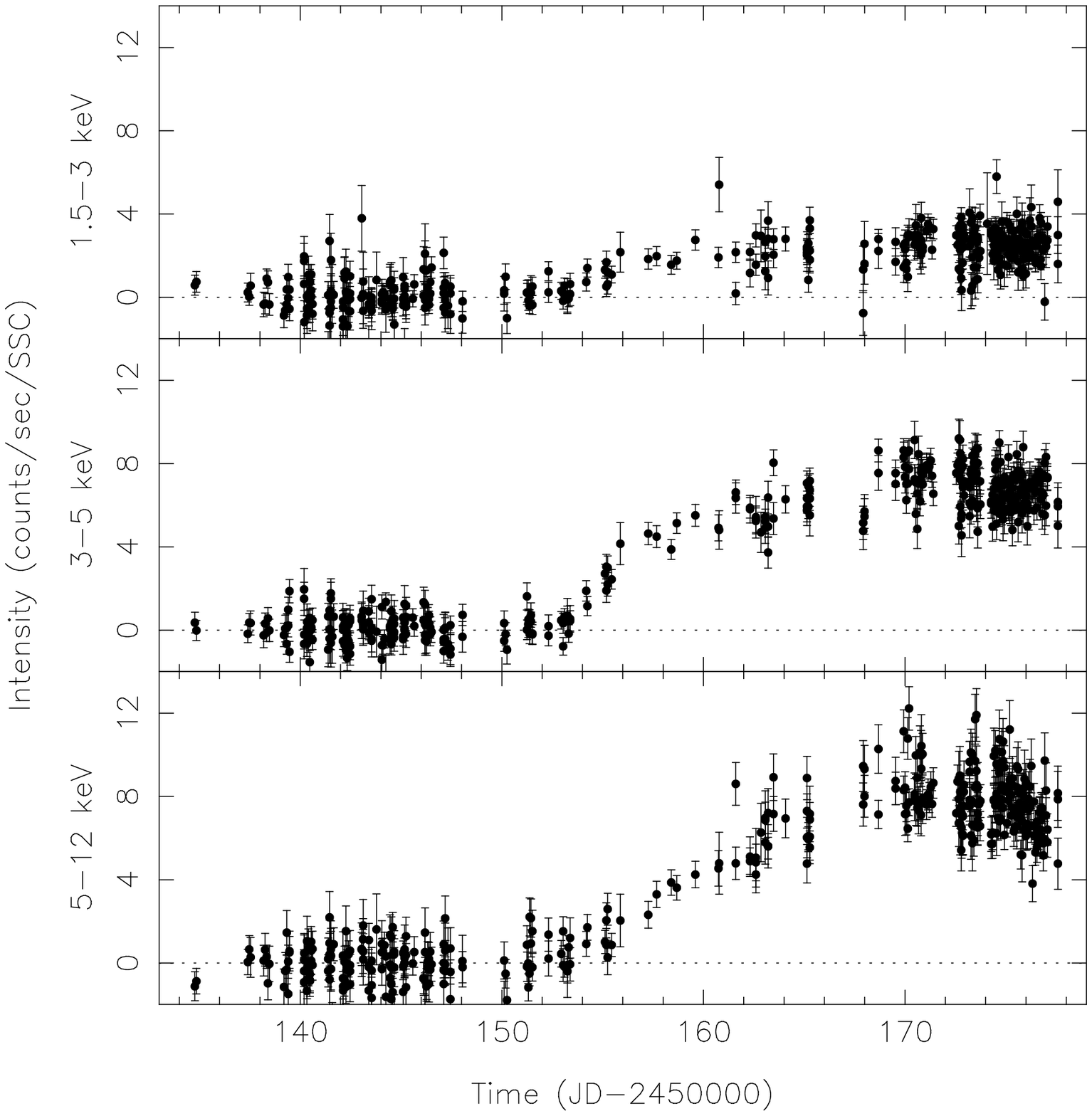,bbllx=65pt,bblly=194pt,bburx=534pt,bbury=739pt,angle=0,height=15cm}}
}
\caption{}
\label{start_1996}
\end{figure}

\begin{figure}
\centerline{\hbox{
\psfig{figure=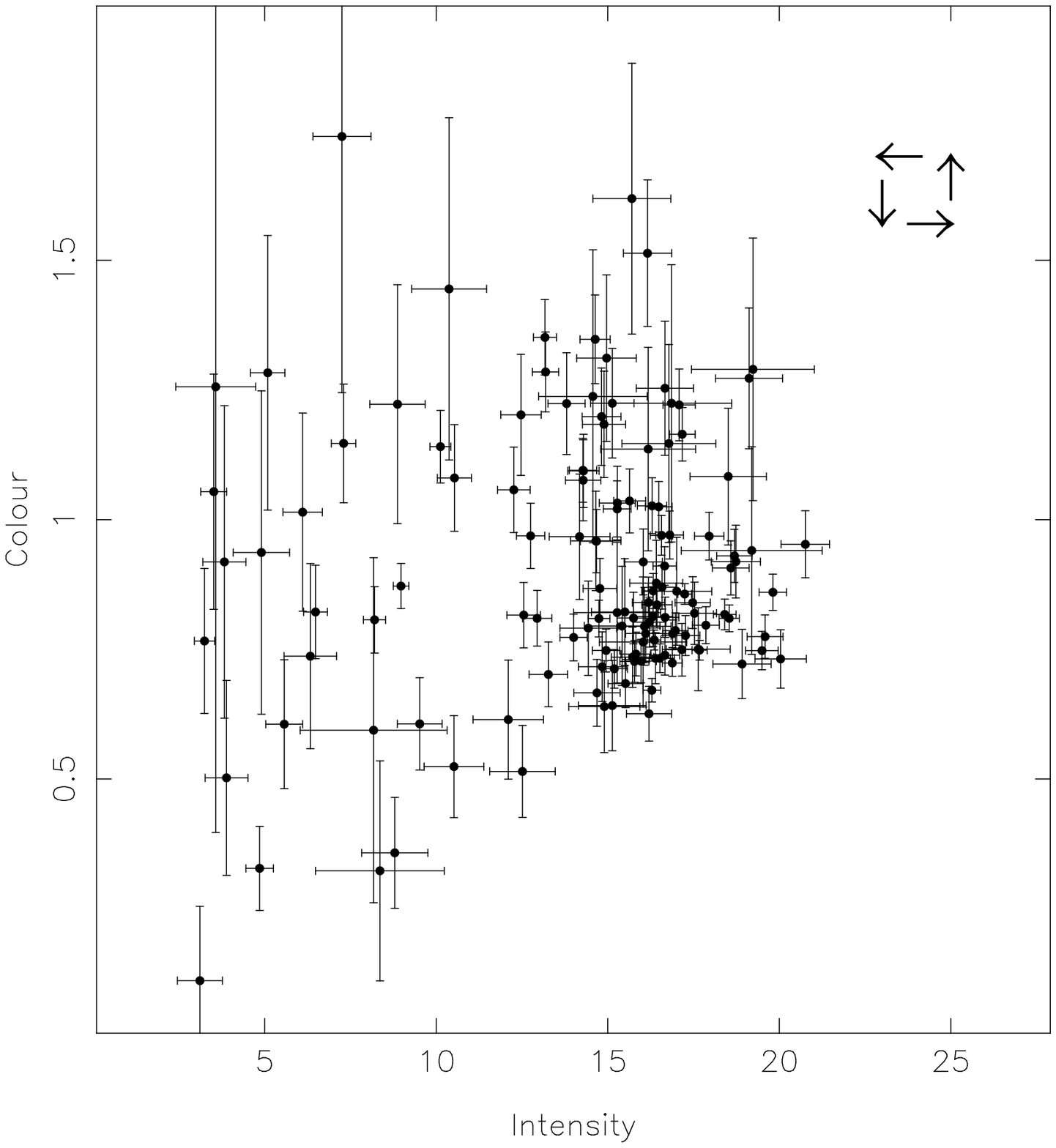,bbllx=66pt,bblly=194pt,bburx=534pt,bbury=739pt,angle=0,height=15cm}}
}
\caption{}
\label{xte_hid}
\end{figure}

\begin{figure}
\centerline{\hbox{
\psfig{figure=kuulkers_1630_fig9.ps,bbllx=65pt,bblly=194pt,bburx=534pt,bbury=595pt,height=13cm}}
}
\caption{}
\label{arvind}
\end{figure}

\end{document}